\documentclass[11pt,twoside]{article}  
\usepackage{apn3conf}

\def\bar{\overline}

\def\bul{{$\bullet$\ }}
\def\part{\partial}

\def\beqn{\begin{eqnarray}}
\def\eeqn{\end{eqnarray}}

\def\ni{\noindent}
\def\.{\mathaccent 95}
\def\beq{\begin{equation}}
\def\ee{\end{equation}}

\def\frac#1#2{{\textstyle{{#1}\over {#2}}}}
\def\ni{\noindent}
\def\lsim{\mathrel{\rlap{\lower4pt\hbox{\hskip1pt$\sim$}}
    \raise1pt\hbox{$<$}}}
\def\gsim{\mathrel{\rlap{\lower4pt\hbox{\hskip1pt$\sim$}}
    \raise1pt\hbox{$>$}}}
\def\sqr#1#2{{\vcenter{\vbox{\hrule height.#2pt
         \hbox{\vrule width.#2pt height#1pt \kern#1pt
         \vrule width.#2pt}
         \hrule height.#2pt}}}}

\def\ctr{\centerline}
\newbox\grsign \setbox\grsign=\hbox{$>$} \newdimen\grdimen \grdimen=\ht\grsign
\newbox\simlessbox \newbox\simgreatbox
\setbox\simgreatbox=\hbox{\raise.5ex\hbox{$>$}\llap
     {\lower.5ex\hbox{$\sim$}}}\ht1=\grdimen\dp1=0pt
\setbox\simlessbox=\hbox{\raise.5ex\hbox{$<$}\llap
     {\lower.5ex\hbox{$\sim$}}}\ht2=\grdimen\dp2=0pt

\def\doublespace {\smallskipamount=6pt plus2pt minus2pt
                  \medskipamount=12pt plus4pt minus4pt
                  \bigskipamount=24pt plus8pt minus8pt
                  \normalbaselineskip=24pt plus0pt minus0pt
                  \normallineskip=2pt
                  \normallineskiplimit=0pt
                  \jot=6pt
                  {\def\smallskip {\vskip\smallskipamount}}
                  {\def\medskip   {\vskip\medskipamount}}
                  {\def\bigskip   {\vskip\bigskipamount}}
                  {\setbox\strutbox=\hbox{\vrule 
                    height17.0pt depth7.0pt width 0pt}}
                  \parskip 12.0pt
                  \normalbaselines}

\font\gkvec=cmmib10                         
\def\bomega{\hbox{{\gkvec\char33}}}                  

\def\lb{\langle}
\def\rb{\rangle}

\def\bw{\bar{\omega}}
\def\bv{\bar V}
\def\bB{\bar B}
\def\ts{\times}
\def\lb{\langle}
\def\rb{\rangle}
\def\curl{\nabla {\ts}}

\def\bfv{{\bf v}}

\def\bfj{{\bf j}}

\def\bfw{{\bomega}}

\def\bfb{{\bf b}}

\def\bbB{\bar{\bf B}}
\def\nb{\nabla}
\def\curl{\nb\ts}

\def\b0{b^{(0)}}
\def\v0{v^{(0)}}
\def\w0{\omega^{(0)}}
\def\bb0{\bfb^{(0)}}
\def\bv0{\bfv^{(0)}}
\def\bw0{\bfw^{(0)}}
\def\bj0{\bfj^{(0)}}

\def\ni{\noindent}

\begin{document}   

%
%
%
%

\title{Dynamo-Driven Outflows in Pre-Planetary Nebulae}

%
%
%

\author{Eric G. Blackman }
\affil{Physics \& Astronomy Dept., Univ. of Rochester, Rochester NY, 14627}

%
%

\contact{ Eric G. Blackman}
\email{blackman@pas.rochester.edu}

%
%
%
%
%

\paindex{Blackman, E.G.}

%
%

\authormark{Blackman}

%
%

\keywords{magnetic fields, bipolar outflows, accretion disks, 
dynamos, multipolar outflows, binary systems}


\begin{abstract}          
The plethora of asymmetric planetary nebulae and the curiously high 
momenta of pre-planetary nebula outflows 
suggest that rotational energy is extracted from
the engines.
Magneto-rotational outflows driven by dynamos might be operating therein.
I summarize scenarios involving
(a) an isolated AGB star and 
(b) a binary.
The efficacy of (a) requires re-establishing differential rotation  
over the AGB star's lifetime, whereas (b) delivers the 
angular momentum at the end of the AGB phase when needed, and  
may involve turbulent disks.  Both 
can produce fields that drive outflows of the required 
mechanical luminosity and 
momentum, though weak points and  open questions require further study.

\end{abstract}


\section{Observations Suggest a Role for  Magneto-Rotational Outflows}


\ni  
Powerful anisotropic outflows appear in a wide range of 
astrophysical sources and magneto-rotationally driven models are often a
favored explanation.  Maybe 
asymmetric outflows (e.g. Balick \& Frank 2003) in planetary nebulae (PPNe)  
and pre-planetary nebulae (PPNe) also involve dynamically important
magnetic fields (Aller 1958). 
Magneto-rotational driving is particularly 
appealing for PPNe winds which are more powerful
than PNe winds:
The  observationally inferred rates of  mechanical momentum and energy deposition for PNe 
are (Kwok 2000)
${\dot \Pi}\sim 10^{27} (M_{pn}/0.04{M_\odot})(v_{pn}/40{\rm km/s})(\Delta t_{PN}/10^4{\rm yr})^{-1}$ erg/cm,
and ${\dot E}\sim 10^{34} ({\dot \Pi}/10^{27}{\rm erg/cm})(v_{pn}/40{\rm km/s})$  
{erg/s}, 
where $M_{pn}$ is the swept up mass in the PN,  
$v_{pn}$ is the speed of this mass, 
and the acceleration time $\Delta t_{PN} \le$ age of the  PN. 
But observations of PPNe (Bujarrabal et al. 2001)
reveal values up to 
${\dot \Pi}\lsim 3 \ts 10^{29}(M_{ppn}/0.5M_\odot)(v_{ppn}/200{\rm km/s})
(\Delta t_{ppn}/10^3{\rm yr})^{-1}$ erg/cm, 
and 
${\dot E}\lsim 10^{37}(\Pi/10^{40}{\rm erg/cm})
(v_{ppn}/200{\rm km/s}){\rm erg/s}$
where the acceleration time $\Delta t_{ppn}<<$ age of the PPN.
The PPNe wind momenta are often $\sim 10^3$
times that carried by radiation  
even when multiple scattering is considered.

A large scale magnetic field can act as a drive belt
that extracts rotational energy to drive a wind
from the engine where the field lines are anchored (e.g. Spruit 1996).
When the engine is a star, the extraction spins the star down.
When the engine is an accretion disk, a fraction
of the accretion luminosity gets redirected into the wind.
In the  Blandford \& Payne (1982) type model, 
mass outflow is driven by the 
toroidal magnetic field pressure gradient along poloidal field lines, 
though in the rotator's co-rotating frame near the footpoints, 
the processes  can be thought of as centrifugal launching along
favorably inclined poloidal field lines.
Consideration of magnetically mediated outflows 
raises natural questions in the PPNe/PNe context:
Where do the required fields come from?
How do these mechanisms work for single star or binary/disk systems?
 Are the outflow powers and momenta consistent with the observations?

\section{ Why Would Dynamos be Needed to Produce the Fields?}

Convection driven and  magneto-rotationally driven MHD turbulence 
(Balbus \& Hawley 1998) are fully developed 
in in AGB stars and in accretion disks (the two 
relevant environments) respectively.
Large scale magnetic fields can then  experience turbulent 
diffusion so intense that the field is not  flux-frozen on the relevant 
dynamical or advection time scales (Lubow et al. 1994; Blackman
\& Tan 2003 (BT)). This highlights the need for  
dynamo amplification. The fields also be large enough to 
avoid diffusing before escaping to the corona where they are needed.
In coronae, the fields can further open up.
Coronal loops which carry magneto-centrifugal winds should be 
at least as large as the radial scale of the anchoring rotator
so that material can  be accelerated to super-Alfv\'enic velocities
before reaching the loop-tops and avoid sliding back down.

\section{Some Recent Developments in Large Scale Dynamo Theory}

Dynamos amplify or sustain magnetic fields by draining energy from 
random and shear motion.
Nonhelical (direct) dynamos amplify fields  at wave numbers $k \ge k_f$ whereas 
helical (inverse) dynamos can also amplify fields at $k < k_f$,
where $k_f$ is the smallest wavenumber
at which the flow is turbulent. 
Here ``helical'' refers to whether the driving turbulence
possesses a non-vanishing pseudoscalar helicity such as 
kinetic helicity $\lb\bfv\cdot\curl \bfv\rb$, where 
$\bfv$ is the turbulent velocity, and the brackets
imply averaging within a hemisphere.
The nonhelical dynamo increases magnetic energy 
as the turbulence stretches 
magnetic field lines in a nearly random walk, 
but it is the  helical dynamo (e.g. ``$\alpha_d\Omega$,'' see Moffatt 1978)
which produce 
the large scale fields needed for outflows.
Helical turbulence can be supplied by
convection in a star, or by the magneto-rotational instability
in a stratified
disk (Brandenburg et al. 1995; Brandenburg 1998; BT).




While 3-D MHD simulations are essential for understanding dynamos,
 they are not always practical for modeling. 
For the latter, mean field theory (e.g. Moffatt 1978) is used.
Here the field is broken into mean and fluctuating components.
The helical dynamo growth coefficients become correlations of fluctuating
quantities, and the mean field evolution is solved for.
Traditional treatments do not dynamically determine the saturation strength of 
the large scale field;  they either take the growth to 
be steady (=kinematic theory), or presume a 
form of  quenching by the growing magnetic field without 
conserving  magnetic helicity, 
a fundamental invariant of ideal MHD. 
A recent theory which alleviates these limitations by including 
the dynamical evolution of the velocity and magnetic helicity 
(Blackman \& Field 2002) is illustrated below.

Fig. 1 below (from Blackman \& Brandenburg 2003) 
shows the helical $\alpha_d\Omega$ dynamo in 
a rotating northern hemisphere.
The kinematic theory is shown in  (a) and (b)
while (c), (d) and (e) show the dynamic theory.
(a): To conserve angular momentum, rising (falling) blobs in a stratified 
medium twist in the opposite (same) sense to the underlying rotation.
This implies a non-vanishing $\lb\bfv\cdot\curl\bfv\rb$, averaged in a given 
hemisphere. This ``$\alpha_d$'' effect implies 
that a large scale toroidal loop in the northern hemisphere 
incurs a right-handed writhe, and a radial field component as it rises.
(b): Differential rotation (the ``$\Omega$'' effect) 
at the base of the loop 
shears the radial component, amplifying the toroidal field.  
Dissipation or ballooning/opening  of the top part of the loop 
allows for a net flux gain through the rectangle.  
(c): Same as (a) but now with the field represented as a ribbon. 
The right-handed writhe of the large scale loop 
is accompanied by a left-handed twist along the tube,
incorporating magnetic  helicity  conservation.  
(d): Same as (b) but with field represented as ribbon/tube.
(e): Top view of the combined twist and writhe
The backreaction force which resists bending 
is the small scale twist.
Diffusing the top part of the loops allows 
a net flux generation in the rectangles of (a)-(d), and helps 
alleviate the backreaction.

\vspace{.1cm}\hbox to \hsize{\hfill\epsfxsize9.5cm
\epsffile{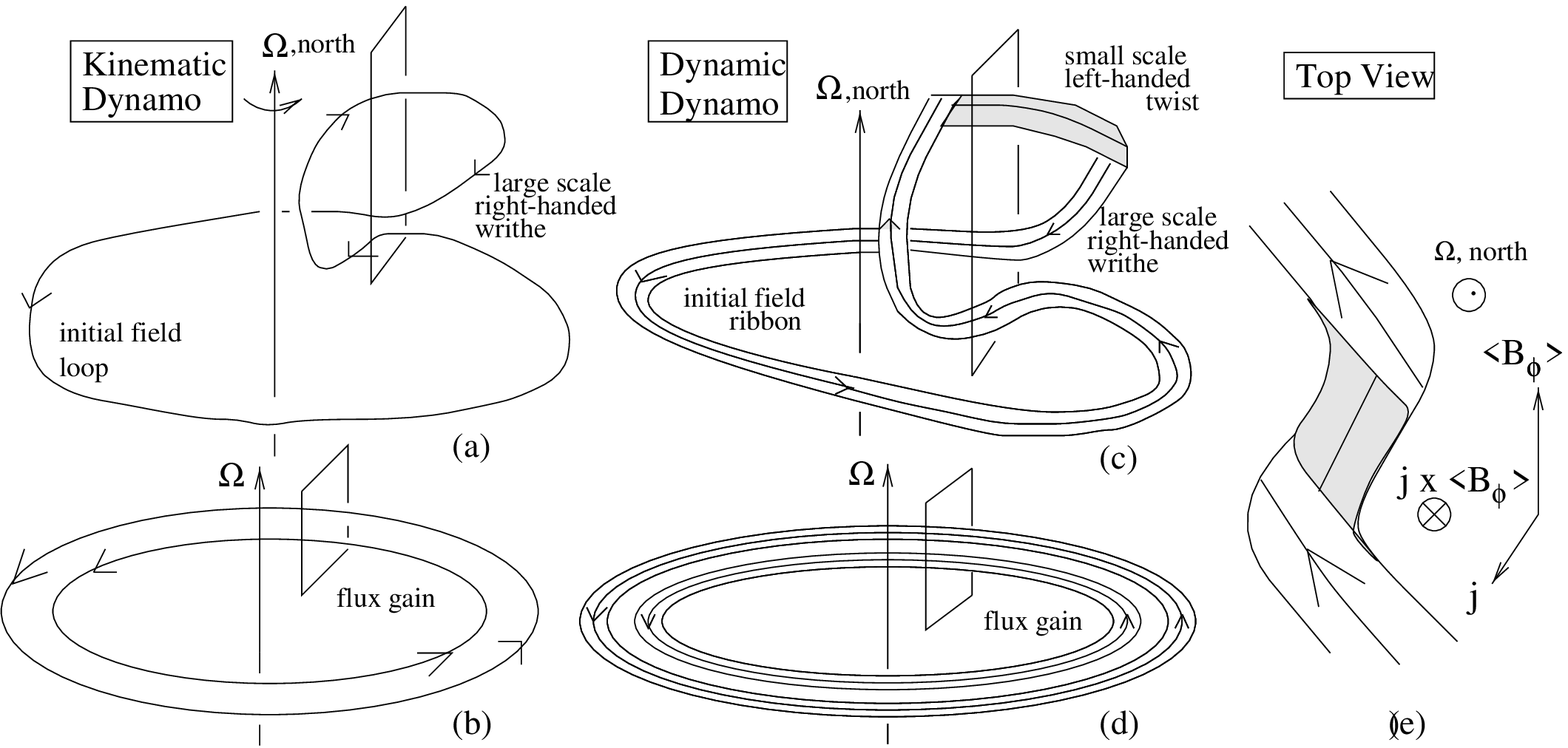} \hfill } 







The dynamic role  of magnetic helicity 
is seen in simulations (Brandenburg 2001)
and  modeled analytically (Field \& Blackman 2002; Blackman \& Field 2002).
Although the small scale magnetic helicity growth 
drives the helical dynamo to a steady state, 
the large scale field grows large enough for most astrophysical applications.
Application of these principles to 
disks and stars 
with boundaries is work in progress. 
In addition, buoyancy rather than kinetic helicity may be the initial
driver of the helical $\alpha_d$ effect 
in disks (Brandenburg 1998).  
This affects the initial sign of $\alpha_d$  in each hemisphere,  
but magnetic helicity evolution remains 
a key to understanding the saturation.

\section{Single Star Scenario: AGB  Dynamo and Magnetic Explosion}

This scenario (Blackman et al. 2001 (BFMTV)) begins with a main sequence (MS) 
star $\sim 3M_\odot$ of rotational speed $\sim 200$ km/s.
At the end the MS lifetime, the core collapses and the envelope 
expands, and angular momentum is assumed to be conserved 
on spherical mass shells. 
The rotation profile satisfies $\Omega\propto 1/r^2$
which is then combined with a numerical AGB stellar model 
(SDK1401, Steve Kawalar, personal communication).
In SDK1401, the convection zone extends from the outer stellar radius 
$5\times 10^{12}$cm down to the core radius $r_c=9\times 10^{10}$cm.
At the ``interface'' between the core and the envelope,  
$\Omega\sim 2 \times 10^{-5}$ cm/s.  

The combination of a strong
differential rotation layer and overlying convection 
zone may lead to an ``interface dynamo'' (e.g. Markiel \& Thomas 1999).
Here the poloidal field is generated by 
a helical $\alpha_d$-effect layer in the convection zone.
Diffusive transport  (or turbulent pumping; Tobias et al. 1998) 
bleeds the poloidal field to the shear layer beneath the convection
zone where it is further amplified by linear stretching
(``$\Omega$'' effect). BFMTV applied the 
code of Markiel \& Thomas (1999) to the  AGB star, 
(The $\alpha_d$ formula used was consistent
with magnetic helicity conservation.)
At $r\simeq r_c$ the saturated field is 
$B\sim 5 \times 10^4$ G.

The scenario requires this dynamo to be 
steady for $\sim 10^6$ yr until the envelope of the AGB star 
is radiatively driven away in the late AGB stages.
Eventually, the density  falls such that the field can no longer be stored: 
Turbulent pumping 
can be gauged by assuming it to be proportional to a 
fixed convective energy density transfer rate $\rho v^3$. 
As $\rho$ decreases, $v$ increases and 
the turbulent pressure, $\propto \rho v^2$,  decreases with decreasing $\rho$, 
eventually dropping  below $B^2/8\pi$ at the core.
At this point, a ``magnetic explosion'' associated with toroidal
pressure could ensue and drive the PPNe outflow.
The magnetic explosion is driven by 
Poynting flux, and surface integration reveals  
a maximum mechanical luminosity (Blackman, Frank, Welch 2001 (BFW))
$
L_{m} \sim 10^{37}
\left({B\over 5\ts 10^4{\rm G}}\right)^2
\left({\Omega_c\over 10^{-5}/{\rm s}}\right) \left({r_c\over r_\odot}\right)^3
{\rm erg/s}, 
$
 where $r_c$ and $\Omega_c$ are
the core radius and angular speed respectively. 
This high $L_m$ lasts for a magnetic  spin down time of the star 
($\sim 100$ yr, starting after the envelope is lost) 
and then decays exponentially.
The asymptotic outflow speed should approach 
$v_{\infty}\sim \Omega_0 r_A$, where
$r_A$ is the Alfv\'en radius, which can be less than or
comparable to the escape speed 
$v_{esc}\sim 520\left({r_c/10^{11}{\rm cm}}\right)^{1/2}
\left({M_*/M_\odot}\right)^{1/2} $ km/s,
depending on $r_A$.  
Matt et al. (2003) (these proceedings) have simulated 
a magnetic explosion. The above results are consistent with 
the requirements of PPNe (see Sec. 1).
(Imai et al. (2002) have observed a collimated AGB jet.)

Unfortunately, the present interface dynamo models, while  
including a saturation of 
the $\alpha_d$ effect, do not include  saturation of 
the $\Omega$ effect.
In reality, amplification of the magnetic field at the shear
layer drains shear energy, and transfers angular momentum from the core to the envelope, slowing
down the core within $\le 100$ yr, 
unless the differential rotation is re-seeded
throughout the $10^6$ yr lifetime of the AGB phase.
This is important because matching the  PPNe observations
requires the field to be anchored in a rapidly rotating 
core. 

The so called ``Lambda effect'' (c.f. R\"udiger 1989) 
might be able to re-seed the differential rotation.
This mechanism involves the interplay between the 
overall rotation and asymmetric convection
to produce a steady-state rotation profile. 
The helical interface dynamo has to be solved with 
a dynamically coupled Lambda effect.
This awaits attention, but note that there is enough energy in the fusion driven  
turbulent convection to sustain the differential rotation:
The turbulent energy density can be calculated from 
from the aforementioned AGB stellar model SDK1401
and the result is (J. Nordhaus personal comm.)
$\sim 5 \ts 10^{43}$ erg. This can be  compared 
with the magnetic field energy in the 
shear layer from BFMTV, which is $E_{m}
\sim 10^{41} $ erg, which in turn is about 1/8 the shear 
energy in the interface layer.
Thus only $\sim 2$\% of the energy from the convection needs to be steadily 
drained into the desired differential rotation.


\section{Binary Scenarios: Disk Dynamos and/or Primary Core Spin-up}

Binary scenarios do not suffer from the sustained shear problem
just discussed because they deliver the angular momentum near the end of 
the AGB phase (e.g. Soker 1998), just when needed 
to drive the outflows. 

Such angular momentum transfer could occur in the late
AGB stage by common envelope (CE) evolution  (e.g. Iben \& Livio 1993), 
as the secondary star penetrates the distended AGB envelope
and transfers orbital angular momentum to it.
The AGB star can shed $80\%$ of its mass within years of the onset of CE,
depending on the binary parameters.
During the shedding, the companion spirals nearer to the core.
While accretion of envelope material 
can occur onto the secondary (e.g. Morris 1987; Soker \& Livio 1994; 
Mastrodemos \& Morris 1998), much larger (needed) accretion rates  
arise when an accretion disk forms around the primary 
via disruption of the secondary (Reyes-Ruiz \& L{\'o}pez 1999).
Favored systems for this mode
of accretion 
involve an evolved AGB star with mass $2.6 \le M_*/M_\odot\le 3.6$,
a  secondary with mass ($\le 0.1 M_\odot$), and the 
initial binary separation $\ge  200 R_\odot$. 
Reyes-Ruiz \& L{\'o}pez (1999) find that the disk accretion rate onto the residual
primary core evolves 
in time (after an initial $\sim 1 $yr viscous adjustment period) 
as $
\dot{M}_d = \dot{M}_{do} (t/{\rm yr})^{-5/4}
{\rm{M}_\odot~{\rm yr}^{-1}}.
$
Typical values of the initial accretion rate  $\dot{M}_{do} \simeq 10^{-3}{\dot M}$/yr.

Once the disk forms, 
a helical disk dynamo 
(e.g. Pudritz 1981; Brandenburg et al. 1995; 
Reyes-Ruiz \& Stepinksi 1995; Campbell 2000; BT)
provides the mean field strengths from which one can determine 
the magnetic luminosity $L_{m}$, available to drive an outflow.
This can be estimated as the integrated Poynting flux flowing 
through the disk surface (or, the Poynting flux through  the Alfv\'en surface 
(BFW; Frank \& Blackman 2003 (FB)).  Tan \& Blackman (2003) find 
$
L_{m} \sim \bB_p\bB_\phi\Omega_i r_i^3 
\sim {\alpha_{ss}^{1\over 2}G{\dot M}_d M_*\over 
r_i}
\sim 10^{37}\left({\alpha_{ss}\over 0.1}\right)^{1\over 2}
\left({M_*\over M_\odot}\right)\left({r_i\over 10^{11}{\rm cm}}\right)^{-1}
\left({t\over 1{\rm yr}}\right)^{{-5\over 4}}
{\rm erg\over s}
$
where $\alpha_{ss}\lsim 0.1$ is the assumed disk viscosity parameter
(Shakura \& Sunyaev 1979), $r_i$ is the inner radius, and
$c_s$ and $\Omega_i$ are the sound and rotation
speeds there.
The ${\dot M}_d$ enters because $\rho \propto  {\dot M}_d$, 
and the mean surface poloidal and toroidal fields satisfy 
${\overline B}_p \sim \rho^{1/2} \alpha_{ss} c_s$ 
and ${\overline B}_\phi \sim \rho^{1/2} \alpha_{ss}^{1/2} c_s$. 
A calculation of 
acceleration along a poloidal field line gives the asymptotic velocity 
$v_{\infty}\sim  \Omega_i r_A \sim 
2^{1/2} \Omega_i r_i
\sim 730 \left(M_*/ M_{\odot}\right)^{1\over 2}\left({r_i/10^{11}{\rm cm}}\right)^{1\over 2}$ km/s 
(FB).  For a rotationally supported  disk, 
the asymptotic wind velocity is always near or greater than the escape
speed. 
The mechanical luminosity and momentum delivered 
are consistent with the  PPNe requirements of Sec. 1.

Rather than form a disk as just described, an alternative is 
that the companion
could spin up both a slowly rotating core and  envelope,
leading to significant differential rotation 
between the two (assuming a turbulent viscosity in the envelope).
This could rejuvenate the interface dynamo of Sec. 4 during CE
in the late AGB stage just when needed. The time scales of spin-orbit
synchronization (Zahn 1977), envelope ejection, and viscous dissipation 
(Iben \& Livio 1993) must be compared, but  
this alternative might mean that a wider range of binary systems 
could produce magneto-rotational outflows.


\section{\bf Open Questions}

Asymmetric outflows in PPNe and PNe 
signature the transport of angular momentum, and   
magnetically mediated outflows may produce the required power and 
momenta. Maser (Miranda et al. 2001; Vlemmings et al. 2003) and core X-ray 
(Kastner et al. 2003) observations loosely support this 
general paradigm. 

However, key unresolved issues remain:
(1) Is there a need for collimated outflows from both the 
AGB star and  disk? Are both self-collimated or
does the latter collimate the former?
What might misaligned nested winds produce (BFW)?
(2) Do all PPN/PNe produce collimated jets when the sources are young? 
(3) Large scale nonlinear dynamo theory 
is in its infancy for realistic boundary terms.
How do helical dynamo fields open up and dynamically relax 
in the coronae?
(4) How far is momentum carried as Poynting flux?
(e.g. hydromagnetic  ``fling'' models vs. magnetically dominated 
``spring'' models, or nested hybrid models? Ustyugova et al. 2000).
(5) MHD jet simulations have been separate from MHD dynamo simulations. 
Simulations capturing  both together have not been
done, though using a dynamo produced field and then simulating
the outflow provides encouraging results (von Rekowski et al. 2003).
Non-steady calculations are important in this context.
(6) The single AGB wind scenario 
will only work for PPNe if convection + rotation
can steadily re-seed the differential rotation
to ensure a large enough 
Poynting flux at the end of the AGB phase.
If not,  binaries may be absolutely required for asymmetric 
PNe and PPNe (e.g. Soker \& Livio 1994; Soker 2001).
Stellar evolution models which include magnetic fields and rotation
are needed.
(7) To what extent are the strong outflows in PPNe correlated with the presence
of binaries?  Binaries with brown dwarfs or planets are not easily detected,
but are favored for accretion around the AGB  core.
(8) How restrictive are 
the initial binary parameters for which common envelope
evolution simulations (Demarco et al. 2003) 
would produce a suitably accreting disk to power PPNe?
Is this relaxed by the scenario at the end of Sec. 5?
(9) Consequences of internal dynamo produced fields
for surface X-ray emission in  AGB stars and disks 
need more study.
(10) Could winds from the primary vs.
secondary disks be distinguished by abundance differences (FB)?

\end{document}